\documentclass[9pt,twocolumn,twoside]{osajnl}

\usepackage{mathptmx}
\usepackage{mathtools}
\usepackage{hyperref}
\usepackage{cleveref}
\usepackage{graphicx}
\usepackage{multirow}
\usepackage{longtable}
\usepackage{booktabs} 
\usepackage{color}
\usepackage{siunitx}

\journal{optica} 

\title{ A $\chi^{(2)}$-based AlGaAs Phase Sensitive Amplifier with Record  Gain, Noise and Sensitivity}

\author[1]{Zhizhong Yan}
\author[1] {Haoyu He}
\author[1] {Han Liu}
\author[1] {M. Iu}
\author[1] {O. Ahmed}
\author[1] {E. Chen}
\author[1] {P. Blakey}
\author[2] {Youichi Akasaka}
\author[2] {T. Ikeuchi}
\author[1,*]{Amr S. Helmy}

\affil[1]{The Edward S. Rogers Sr. Department of Electrical and Computer Engineering, University of Toronto, 10 King’s College Road, Toronto, Ontario, Canada M5S 3G4}
\affil[2]{Fujitsu Laboratories of America, 2801 Telecom Pusingarkway, Richardson TX 75082 USA}

\affil[*]{Corresponding author: a.helmy@utoronto.ca}

\dates{Compiled \today}

\ociscodes{(250.4745) Optical processing devices;  (320.7080) Ultrafast devices.}

\doi{\url{http://dx.doi.org/10.1364/optica.XX.XXXXXX}}

\begin{abstract}
	Phase sensitive amplifiers (PSAs) have the potential to empower substantial advances in emerging generations of optical communication systems as well as classical and quantum on-chip signal processing. While the second order nonlinearity ($\chi^{2}$) is stronger than the third order nonlinearity ($\chi^{3}$), it seldom utilized in semiconductors to realise PSAs owing to the challenges of effectively phase matching the interacting waves as well as countering the two-photon absorption of the pump. In this work, we demonstrate the successful design, fabrication, and characterization of the first $\chi^{2}$-based semiconductor PSA using an efficient phase matching approach and a pulsed pump, in an AlGaAs Bragg reflection waveguides. The reported AlGaAs PSA achieves on chip in-phase gain approaching 30 dB, with a sensitivity of 0.005 photons per pulse. Its performance also approaches the theoretical minimal noise figure (NF) of 0~dB. With such performance metrics and its capability to operate in single mode regime, this PSA could usher in a new era of on-chip quantum circuits.
	
\end{abstract}	

\setboolean{displaycopyright}{true}

\begin{document}
	
\maketitle

\par  \textcolor{black}{Phase sensitive amplifiers} have attracted significant attention due to their ability to mitigate the noise limitations imposed by conventional amplifiers \cite{Tong2011}. Other advantages of PSAs include optical phase and amplitude regeneration, dispersion compensation, suppression of modulation instability, enhanced imaging, and sensing performance through noise suppression \cite{Choi1999a, Mosset2005}. Four-wave mixing (FWM) in fibers and on-chip is the predominant technique for implementing PSAs and uses the third-order optical nonlinearity ($\chi^{3}$) \cite{Cestier2012,Li2015a,Agarwal2014}. However, PSAs have also recently been implemented using second-order ($\chi^{2}$) nonlinearity in \textcolor{black}{periodic poled lithium niobate} (PPLN) waveguides \cite{Umeki2015,Kashiwazaki2019}. The use of $\chi^{2}$ instead of $\chi^{3}$ provides numerous benefits for PSAs, including immunity to stimulated Brillouin scattering, a larger nonlinear coefficient, more effective pump filtering, low spontaneous emission noise, low cross talk, and no intrinsic frequency chirp \cite{Han2010}.
 
\begin{figure*}
	\begin{minipage}{\textwidth}
		\centering
		\includegraphics[width=0.7 \columnwidth]{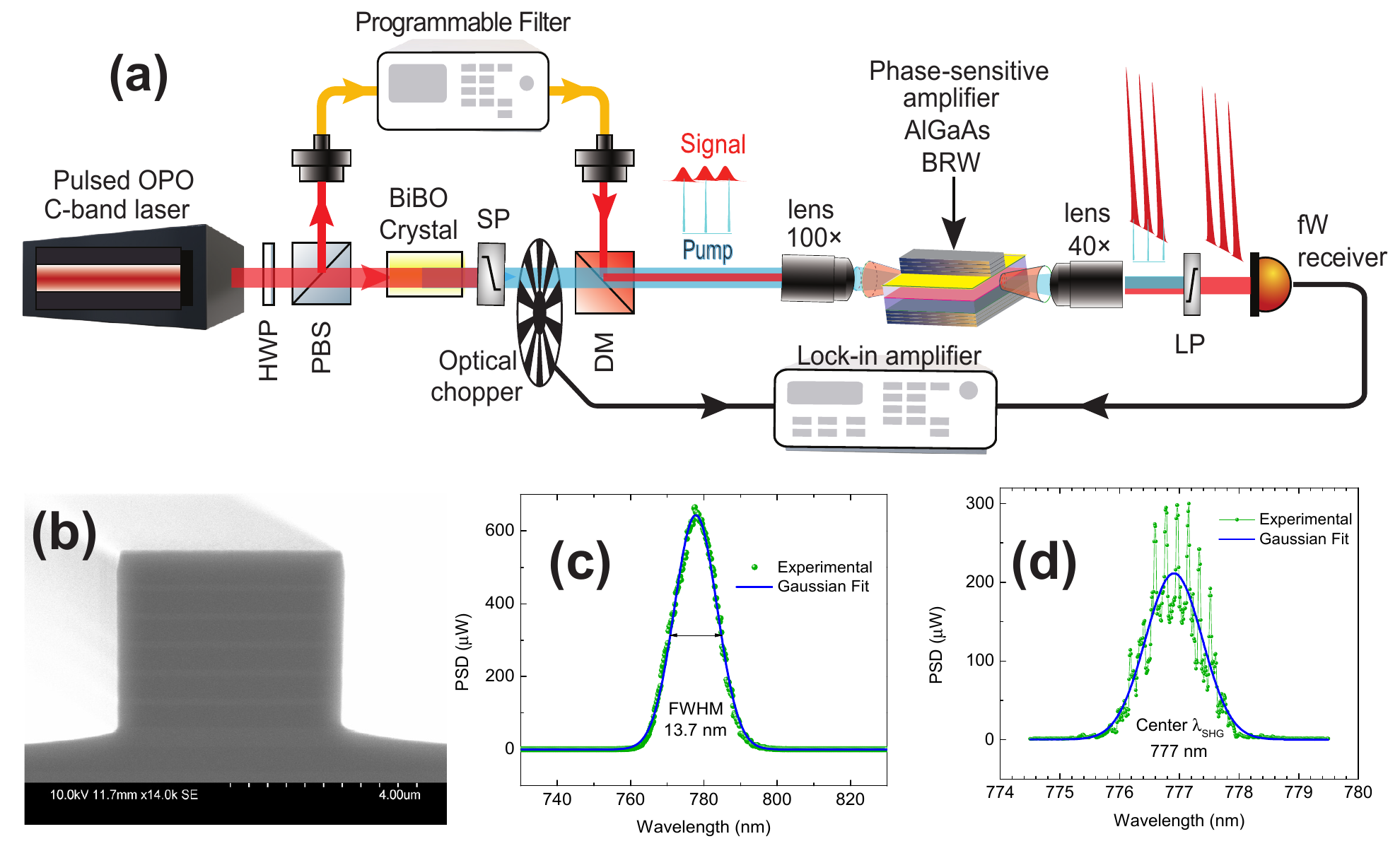}
		\caption{\textbf{AlGaAs PSA chip and experimental setup.} 
			 (a)~Experimental setup. An OPO provides 1560~nm femtosecond pulses that are coherently split between two paths. One path pumps a nonlinear crystal (1~mm thick BiBO) to generate a 777.8~nm pump via SHG. The other OPO path couples into the fiber port of an optical programmable filter (Finisar Model 1000A) that tailors the spectrum into a 0.2~nm FWHM degenerate signal/idler beam whose phase is varied relative to the pump. A dichroic mirror (DM) combines the pump and signal beams for injection into the AlGaAs optical parametric amplifier chip, which is operated as a PSA. The input beam is coupled to a core layer between two stacks of Bragg reflection mirrors. The pump is in the TM mode, while the signal is in the TE mode. The PSA output is long-pass filtered, followed by a PBS cube (now shown) to remove the unwanted TM component of the amplified signal. The output of the PBS is then fed to a FemtoWatt receiver connected to a lock-in amplifier, with an optical chopper in the pump beam path providing a frequency reference. (b)~SEM image of the AlGaAs waveguide profile showing physical dimensions. (c)~Spectrum of the 777.8~nm pump beam generated via SHG by passing the OPO beam through the BiBO crystal. (d)~SHG spectrum of the PSA waveguide, obtained by directly coupling the OPO beam into AlGaAs waveguide input facet.			 
		}\label{fig:PSA_Setup}
	\end{minipage}
\end{figure*}

\par The performance of the AlGaAs PSA chip was characterized through measurements of the in-phase and out-of-phase gain, the phase-locked and phase unlocked noise figures, as well as the sensitivity. The PSAs described here are based on AlGaAs waveguides \cite{Helmy2007} whose typical structure is shown in the experimental setup illustrated in Fig.~\ref{fig:PSA_Setup}~(a). A scanning electron microscope image (SEM) of a typical fabricated PSA device is shown in Fig.\ref{fig:PSA_Setup}~(b). With this architecture, (see Supplementary Information Sec.~1.1), we are able to utilize the second order nonlinearities with record efficiency \cite{Abolghasem2009}. This remarkable efficiency has been confirmed in several difference frequency generation (DFG) \cite{Han2010a} and sum frequency generation (SFG) \cite{Han2009} experiments in our \textcolor{black}{Bragg Reflection Waveguide} (BRW) waveguides. In the pulsed pump configuration, one can also expect the significant effective length reduction with respect to the device physical length due to the reduction in the parametric interaction with length \textcolor{black}{due} to several impairments as will be discussed later. For example, this 1mm long BRW waveguide exhibits near 50~$\mu$m effective length at average pump power above 15~mW as we shall show further in the text.

The pump beam, which had a central wavelength of 777.8~nm, was generated via second harmonic generation (SHG) in a BiBO crystal driven by an optical parametric oscillator (OPO) source lasing at double the wavelength. The pump, at 777.8~nm, had a spectral full width at half maximum (FWHM) of 13.7~nm (6.88~THz) as shown in Fig.\ref{fig:PSA_Setup}~(c), corresponding to a 64~femtosecond pulse width. To determine the PSA's phase-matched pump wavelength, the nonlinear waveguide was first injected using the OPO only (bypassing the BiBO crystal) to obtain its SHG spectrum as shown in Fig.~\ref{fig:PSA_Setup}~(d).   The degenerate SHG wavelength and thus optimally phase-matched pump wavelength was found to be 777~nm for the device described here.

\par Our experimental setup displayed in Fig.~\ref{fig:PSA_Setup} is to measure the PSA phase-dependent gain and noise performances. Conventional methods for gain determination through comparing input and output power present an inaccurate approach for this device, partly because the amount of pump power coupled into the desired mode is challenging to determine accurately due to the multi-mode nature of these structures (Bragg and total internal reflection (TIR) modes). 

\par However, the phase matching condition, which governs the interacting modes based on conservation laws, offers confidence that the gain measured is that for the mode tested at the output, as such this helps exclude other spatial modes and offers amplification only to the single spatial mode, of the parametric light sought.  
	
\par It is important to note that the pump pulse width $ \tau $ is less than $ 100\times 10^{-15} $~second; while we have computed the phase matching bandwidth (BW) to be 1.2~THz. As such the product of the $ \tau \cdot \text{BW} \ll 1 $, which suggests that a single temporal mode approximation is a valid one for the system we study here. Consequently, we are able to develop our single mode PSA characterization model based on a squeezed coherent state acting as the output signal.

As developed in previous work, PSA outputs usually exhibit squeezing in proportion to the amplifier gain, while the spontaneous emission arising from squeezed vacuum contributes background noise \cite{Smithey1992}. Such a description affords a route for the characterization approach to obtaining the PSA gain. This enables the determination of an accurate gain value that does not rely on the knowledge of the in-coupled power \cite{Koashi1993, Slusher1987, Wasilewski2006}. 
\begin{figure}[t!]
	\centering
	\includegraphics[width=0.9 \columnwidth]{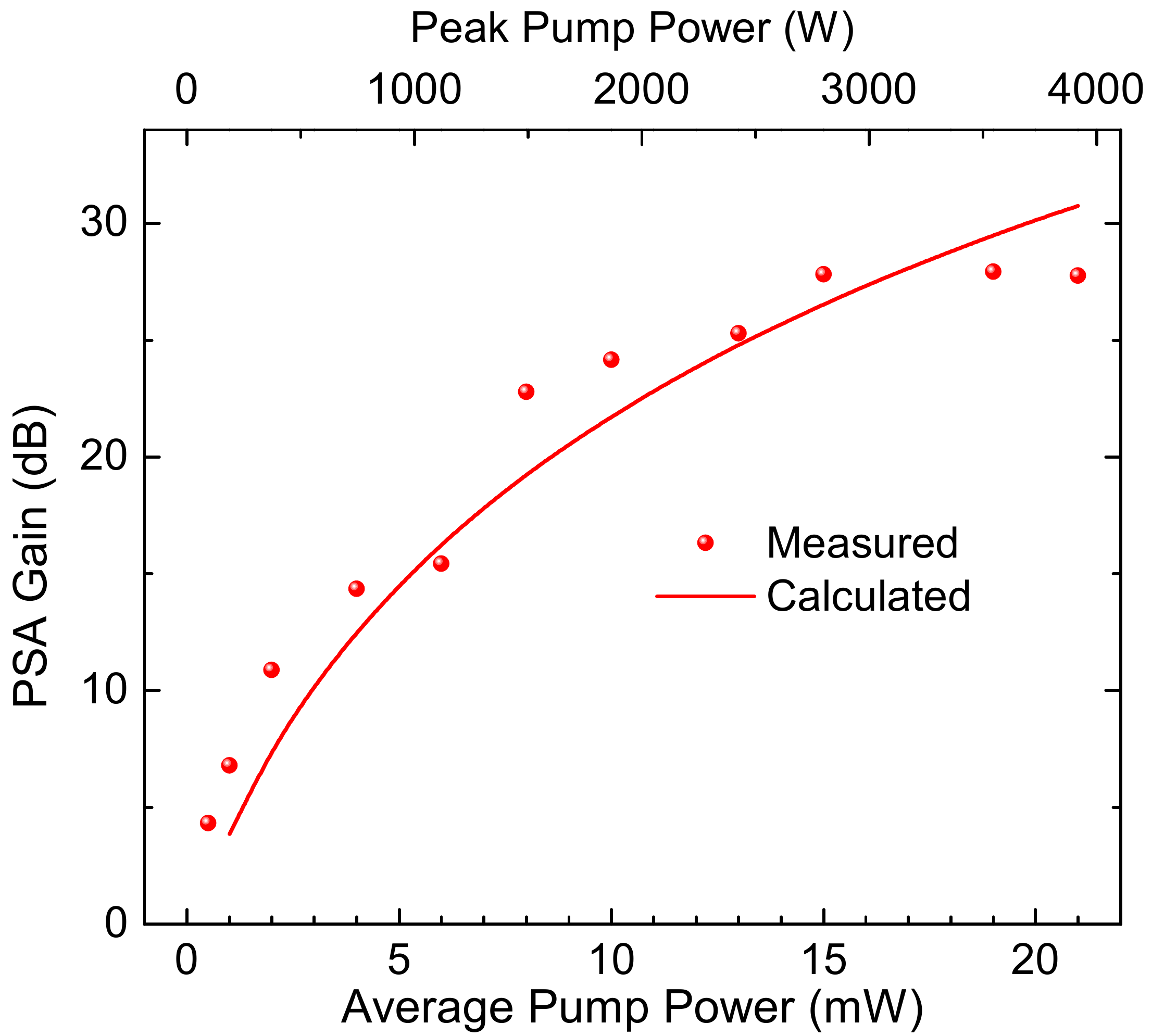}\\
	\includegraphics[width=0.9 \columnwidth]{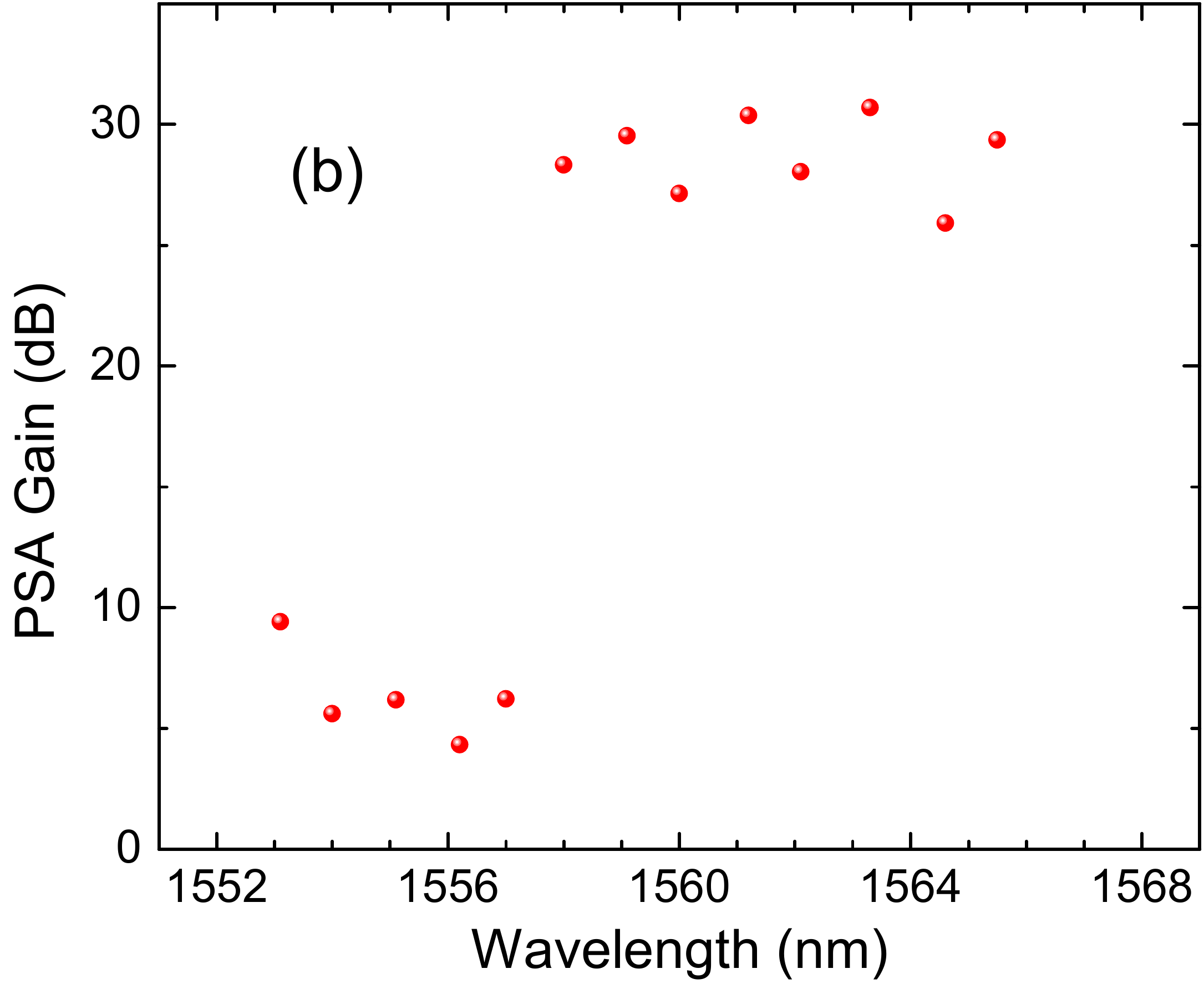}
	\caption{\textbf{PSA experimental gain measurements and simulation results.}  (a) The PSA in-phase gain measured at a fixed wavelength of $1559.7~\pm0.1$ nm as a function of external pump power. The solid red line shows the predictions of the NLSE numerical simulation, where an effective nonlinearity of 70~pm/V was used to fit the measured data. (b) The signal wavelength is swept in 1~nm steps from 1553.1~nm to 1565.6~nm while the external pump power was maintained at 20~mW. Refer to Supplementary Information Sec.~1.2 for details of the PSA gain measurement. }\label{fig:PSAGain}
\end{figure}

\par Using the aforementioned technique which is detailed in  Supplementary Information Sec.~1.2 and 1.4, the dependence of the PSA gain on pump power was investigated, while the signal wavelength and injected signal power both were kept constant. The signal wavelength was $ 1560.0 \pm 0.1 $~nm. The signal power was kept low, around 50~nW, before the 100x input microscope objective, to allow for the largest dynamic range of the detector/receiver. As can be seen in Fig.~\ref{fig:PSAGain}~(a), the measured PSA gain is plotted as a function of the pump power. At average pump powers approaching 20 mW the PSA gain is seen gradually saturating and exhibits an in-phase gain of $\approx $~30~dB. To confirm the role of $\chi^{2}$ in the gain, a linear relationship between the pumped power and the spontaneous emission at the signal wavelength for pump powers from 0.6~mW to 21~mW in confirmed (see Supplementary Information Sec.~1.4). 

The dependence of PSA gain on signal wavelength was also investigated and is shown in Fig.~\ref{fig:PSAGain}~(b). The measurements were carried out using a fixed average external pump power of 20~mW. The plot shows what may be seen as two in-phase gain regimes: a low gain regime when $ \lambda_{s} < 1557 $~nm; and a high gain regime $ g_{PSA} \approx $~30~dB when $ \lambda_{s} \ge 1558 $~nm. This gain profile is determined by the phase matching response, and in Supplementary \textcolor{black}{Information} Sec. 2 we show that this transition between regimes agrees with theory and occurs at the signal wavelength where the phase matching condition is satisfied. Remarkably, our maximum tenable in-phase gain $\approx $~30~dB is achieved within a fraction of  the physical device length of only 1~mm\cite{Tong2011} (see Supplementary Information Sec.~1.4) for phase stability.  
\begin{figure}[t]
	\centering
	\includegraphics[width=0.9 \columnwidth]{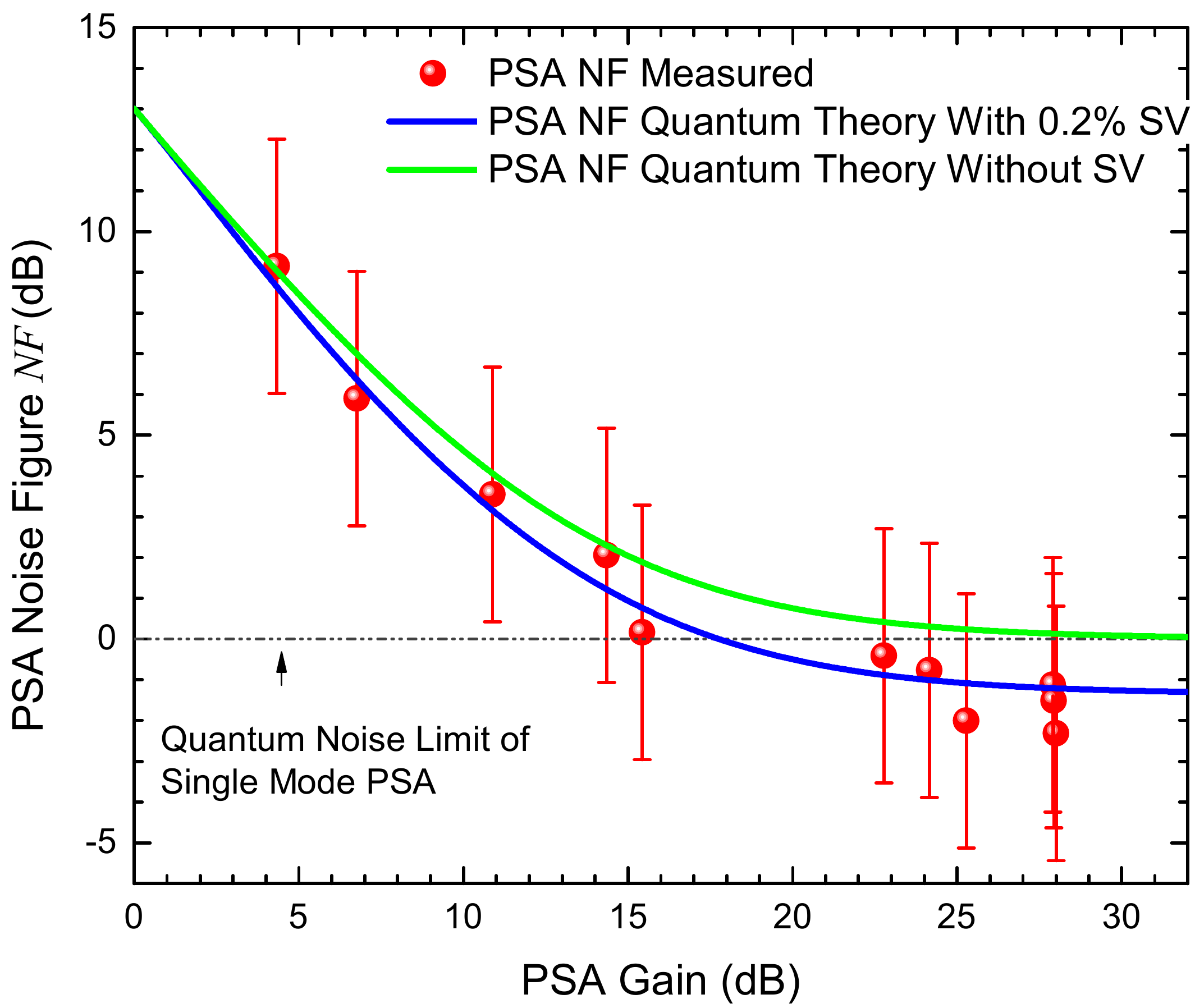}
	\caption{\textbf{Noise Figure measurements.} The noise figure (NF) as a function of amplifier gain was measured for PSA. 
		Noise figure comparison between the experimental measurement and the theory. In the theory calculation, two approaches are displayed: i.e. when the squeezed vacuum is completely removed by the two step experiment (green solid line); 99.8\% squeezed vacuum (SV) photons have been removed, but leave a 0.2\% of residue (blue solid line).
		}\label{fig:noise}
\end{figure}

\par To determine the origin of the large PSA gain observed, we used experimentally-measured device parameters \cite{Han2010} to predict the gain behaviour via coupled Nonlinear Schr\"{o}dinger Equations (NLSEs). The set of NLSEs were solved numerically and found to predict gain performance in good agreement with the experimental results measured, as seen in Fig.~\ref{fig:PSAGain}(a). From the NLSE simulation results, high PSA gain is achieved owing to the large effective nonlinearity (70 pm/V) attainable using this phase matching technique in semiconductors, coupled with the short fs-pulsed nature of the pump. The femtosecond pumping yields substantial peak powers, leading to a large gain even in the presence of significant pump two-photon absorption (TPA) within the PSA. More details are discussed in Supplementary \textcolor{black}{Information} Sec. ~2. 

\par The noise performance of the PSA was measured using the phase sensitive detection (PSD) method \cite{Breitenbach1997} with the aid of a lock-in amplifier, wherein the noise of the squeezed signal with respect to the quadrature phase is characterized. The noise figure (NF) is the most commonly-used 
figure of merit for characterizing the noise added by an amplifier~\cite{Baney2000}. In the PSD method, the amplifier input is injected with a near-ideal shot-noise-limited coherent source, which serves as a probe. The NF is then obtained by comparing the signal-to-noise ratios (SNR) at the amplifier input and output ports respectively \cite{Movassaghi1998}. 

\par  To account for the low signal powers, a quantum optics noise model is used. In this model, output SNR ($ \text{SNR}_{out} $) and the PSA output noise factor $\text{NF}_{\text{PSA}}$ in the presence of optical loss can be obtain as a function of the PSA gain $ g_{\text{PSA}} $. An optical efficiency, $ \eta $, was found to be 4.8\%. The specifics of this model can be found in the Supplementary Information Sec.~3 along with the details for determining the optical efficiency which are listed in Table 1. 

\par To better provide a calibrated benchmark, the $\text{NF}_{\text{PSA}}$ of our device was measured in the PSA configuration. Fig.~\ref{fig:noise} shows the measured PSA noise figure in dB as a function of PSA gain, alongside theoretical predictions under two levels of inclusion of squeezed vacuum contributions. A reference line at 0~dB, indicating the PSA ideal quantum noise limit, has been added to benchmark the PSA performance \cite{Tong2011, Agarwal2014, McKinstrie2005}. The value of $\text{NF}_{\text{PSA}}$ for the PSA should asymptotically approach the intrinsic noise figure value $\text{NF}_\text{PSA}$ that resides well below the quantum noise limit. Our observations are in good agreement for lower PSA gain, but deviates into negative noise figure at high PSA in-phase gain regimes.

\par In Fig. \ref{fig:noise}, the green line shows the theoretical PSA noise figure of the amplified signal as a function of PSA gain and for $\eta=4.8~\%$ which should asymptotically reach a noise figure of 0~dB at high PSA gain. Although, the measured noise figure does show negative values, we attribute this deviation and negative noise figure to the fact that it is technically challenging to completely remove the squeezed vacuum component in practice. In particular, the small PSA signal, $ |\alpha|^2=2.116 \times 10^{-3} $, makes the spontaneous emission term non-negligible. Our model shows that the inclusion of only 0.2\% of squeezed vacuum contributions in the noise figure produces a best fit line, the blue line in Fig.~\ref{fig:noise}, and this non-ideal filtering accounts for this deviation from the theory.

\begin{figure}[t]
	\centering
	\includegraphics[width=0.9 \columnwidth]{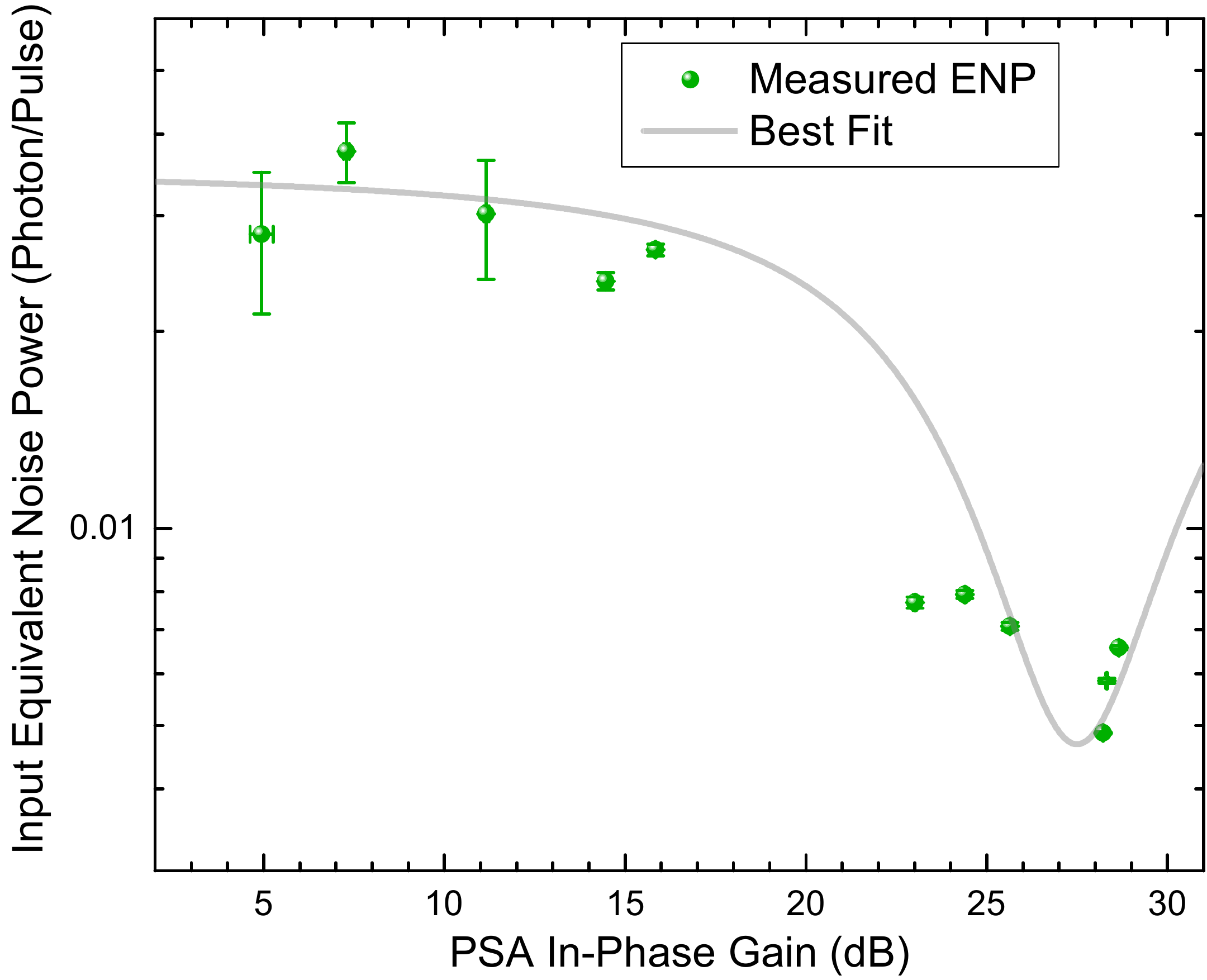}
	\caption{ \textbf{PSA input equivalent noise power (ENP)}. For convenience the ENP has been converted from $ \text{W Hz}^{-1} $ into number of noise photons per pulse, given the input optical photon flux averaged at 0.2 photon per pulse with an energy of 0.80 eV at TE polarization. The optical pulse repetition rate is at 80 MHz in the measurement. The solid gray line is a best-fit to the ENP data.
	}\label{fig:sensitivity}
\end{figure}

\par Enhancing the intrinsic sensitivity of the PSA is of paramount importance for numerous applications including optical communications and quantum information processing (QIP) \cite{Agarwal2014}. The intrinsic sensitivity can be represented by the input equivalent noise power (ENP), which is defined as the minimum signal power that makes the PSA $ \text{SNR}_{out} =1 $ in a 1~Hz bandwidth. In the case of $ \chi^3 $-based PSAs, noise arises from both the nonlinear parametric process and the spectral tails of a strong pump that can be challenging to selectively filter due to the pump's close spectral proximity to the signal/idler. In contrast, for $ \chi^2 $-based PSAs the noise signals that limit sensitivity arise predominately from the nonlinear parametric process, as the pump is much further away in wavelength. This noise source for a $ \chi^2 $ PSA is essentially the spontaneous parametric emission behaving as a random signal superimposed on the input signal. The spontaneous emission power ($ P_{sp} \text{Hz}^{-1} $) is therefore the main limiting factor dictating the sensitivity of such PSAs. To quantify the PSA sensitivity, the spontaneous emission power was measured using the lock-in-amplifier set to a 1~Hz bandwidth. The value of ENP is obtained by dividing $ P_{sp} \text{Hz}^{-1} $ by the PSA in-phase gain. The results are shown in Fig.~\ref{fig:sensitivity}, which plots the ENP in terms of input photons per pulse, as a function of the PSA in-phase gain. The ENP decreases from 0.04 photons/pulse at 7.3~dB PSA gain to the minimum value of 0.005 photons/pulse at 28~dB PSA gain. 

\par  We reported three approaches to study the phase dependence, namely phase dependent gain swing, NLSE numerical simulation and noise figure measurement. All three independent studies agree with one another. 
We therefore have demonstrated a $\chi^{(2)}$ PSA exhibiting extraordinarily high gain ($> 30~\text{dB}$) within an effective length below fractions of 1~millimeter physical length. The PSA gain is highly occurring in the near singular Schmidt mode (Supplementary Information Sec. 1.3). We also found a near ideal noise figure and unprecedented high sensitivity (0.005 photons/pulse) in a compact AlGaAs architecture amenable to standard semiconductor fabrication techniques.

The \textcolor{black}{similar} PSA gain in waveguides has been recently obtained in PPLN \cite{Kashiwazaki2019}. \textcolor{black}{ Moreover, the hetero-integration of PPLN to a SiO$_2$ substrate also allows tighter mode confinement \cite{Pohl2020, Wang2017a}}. However the \textcolor{black}{relative} low-index contrast of PPLN prevents it from \textcolor{black}{directly integrating to existing silicon-like} semiconductor systems. 
 
\par The low noise of the PSA  stage itself affords the opportunity for the entire PSA subsystem to achieve a total noise figure of $\text{NF} << 3~\text{dB}$, which is the limit $\text{NF}$ for phase insensitive amplifiers. This can transform the capabilities of emerging optical communication links.
In particular, the unprecedented sensitivity to sub-single-photon inputs could have profound implications for using PSAs in both quantum key distribution and in  discrete-variable quantum information processing. Recently a 5 km long fiber ($\chi^{(3)}$ PSA) was used in Ref.~\cite{Agarwal2014}, whose estimated distributed amplification sensitivity is around 1 photon per pulse inferred from their spontaneous photon pair measurement. The distributed gain \cite{Agarwal2014, McKinstrie2005} reduces the effects of loss on quantum state quality and thereby mitigates decoherence which severely limits the tenable key rates and processing power \cite{Dailey2015}.

\textcolor{black}{In an optimal case with no restriction on resources, we would be able to achieve coupling loss < 1 dB for the interacting waves at the facets.}  The performance and form factor of our device holds the potential to usher in a new era in integrated PSA subsystems, where the pump, the PSA, and the injection locking subsystem required to lock the incoming signal from a telecommunications link can all be integrated on the same compact platform. 
\newline
\par \textbf{Acknowledgments}\\
The authors would like to thank Z. M. L\'{e}ger for his comments.\\

\par \textbf{\textcolor{black}{Funding}}\\
Natural Sciences and Engineering Research Council of Canada (NSERC)\\

\par \textbf{\textcolor{black}{Disclosures}}\\
The authors declare no conflicts of interest.\\

\par \textbf{\textcolor{black}{Data Availability}}\\
Data underlying the results presented in this paper are not publicly available at this time but may be obtained from the authors upon reasonable request.

\end{document}